# Rayleigh Wave Suppression in Al$_{0.6}$Sc$_{0.4}$N-on-SiC Resonators

Marco Liffredo, *Student member, IEEE*, Silvan Stettler, Federico Peretti, and Luis Guillermo Villanueva, *Member, IEEE*

*Abstract*—We report on the fabrication of a Hybrid SAW/BAW resonator made of a thin layer of Sc-doped AlN (AlScN) with a Sc concentration of 40 at% on a 4H-SiC substrate. A Sezawa mode, excited by a vertical electric field, exploits the d$_{31}$ piezoelectric coefficient to propagate a longitudinal acoustic wave in the AlScN. The resonant frequency is determined via the pitch in the interdigitated transducer (IDT) defined by Deep Ultraviolet (DUV) lithography. The resonant mode travels in the piezoelectric layer without leaking in the substrate thanks to the mismatch in acoustic phase velocities between the piezoelectric and substrate materials. We show the impact of the piezoelectric and IDT layers' thickness on the two found modes. Importantly, we show how thin piezoelectric and electrode layers effectively suppress the Rayleigh mode. While some challenges in the deposition of AlScN remain towards a large coupling coefficient $k_{eff}^2$, We show how wave confinement in the IDT obtains a good quality factor. We also show how modifying the IDT reflectivity allows us to engineer a stopband to prevent unwanted modes from being excited between resonance and antiresonance frequencies. Finally, we validate the simulation with fabricated and measured devices and present possible improvements to this resonator architecture.

## I. Introduction

WITH the recent rollout of 5G and the development of future generations of telecommunications, larger power densities will be required, adding power handling and heat dissipation in the design space for acoustic filters. High powers might cause thermal drift and non-linearity in suspended resonators because the generated heat can only be dissipated through their thin anchors. On the other hand, unsuspended devices allow for efficient heat dissipation due to the large contact area with the substrate. The simplest solution is to use Surface Acoustic Waves (SAW) resonators, but the frequency of operation is limited by energy leaking into the substrate. An improvement over standard SAW are Incredible High Performance SAW (IHP-SAW) Resonators [1]. For this type of resonator, a SAW is confined on the surface of a substrate using a high acoustic velocity layer, which confines the mechanical wave and prevents it from leaking into the substrate. For higher operating frequencies, when Bulk Acoustic Wave (BAW) resonators are a standard solution, solidly mounted resonators (SMRs) are a type of Film Bulk Acoustic wave Resonators (FBARs) where an acoustic wave is confined in a piezoelectric film using an acoustic Bragg reflector layer made of alternating films of high and low acoustic impedance materials. In 2013, an alternative to these two approaches was described. as "A third type of FBAR" [2], also called a hybrid SAW/BAW Resonator (HSB). Successive work showed the feasibility of such a resonator in AlN [3], [4] and AlN/GaN stacks [5].

## II. Design

The originally proposed hybrid SAW/BAW structure of [2] and [6] uses piezoelectric pillars to excite different acoustic waves. FEM simulations typically show two main modes, one which travels mostly on the non-piezoelectric substrate, and a second one that again propagates at the interface, but at a higher frequency. For the resonator described in this paper, we do not define any pillars, just leaving the IDT on top of the AlScN layer. In this case, the two modes mentioned in the previous paragraph turn into a Rayleigh mode (Fig. 1.a, lower frequency) and a Sezawa mode [7] (Fig. 1.b, higher frequency).

The piezoelectric material we use is 40% Sc-doped AlN (AlScN), which provides two benefits for our structure. On the one hand, it has a larger electromechanical coupling than standard AlN, thus enabling wider band filters. On the other hand, the speed of sound is lower than in the case of AlN ($v_l \sim 9000\frac{m}{s}$ [8]) which, even though it results in a lower frequency for a given geometry, it helps confining the energy within the piezoelectric layer ([6]). Indeed, to confine the acoustic energy, the longitudinal wave velocity in the piezoelectric layer must be slower than *any* radiating bulk wave in the substrate. After careful exploration of different substrate candidates ([9], [10], [11], [12], [13]), Only in diamond and 4H-SiC the slowest bulk shear waves are faster than the longitudinal wave in AlScN [6]. Importantly, none of the materials allows for efficient wave confinement for standard AlN. Due to the easier sourcing of 4-inch wafers, we choose SiC as the substrate material, where $v_{sh} \sim 7000\frac{m}{s}$. A very similar structure to the one we propose has been recently reported [14]. Our approach comprises the use of a floating bottom electrode to excite the piezoelectric layer with a purely vertical electric field [15].

Interestingly, by reducing the thickness of the piezoelectric layer (Fig. 1.c) and/or the top Al electrode (Fig. 1.d), the Rayleigh mode can be suppressed while maintaining a good coupling for the Sezawa one. Since the latter has a higher frequency, one must pay attention to the mode dispersion so that the wave velocity is lower than the bulk shear in SiC to avoid bulk radiation.

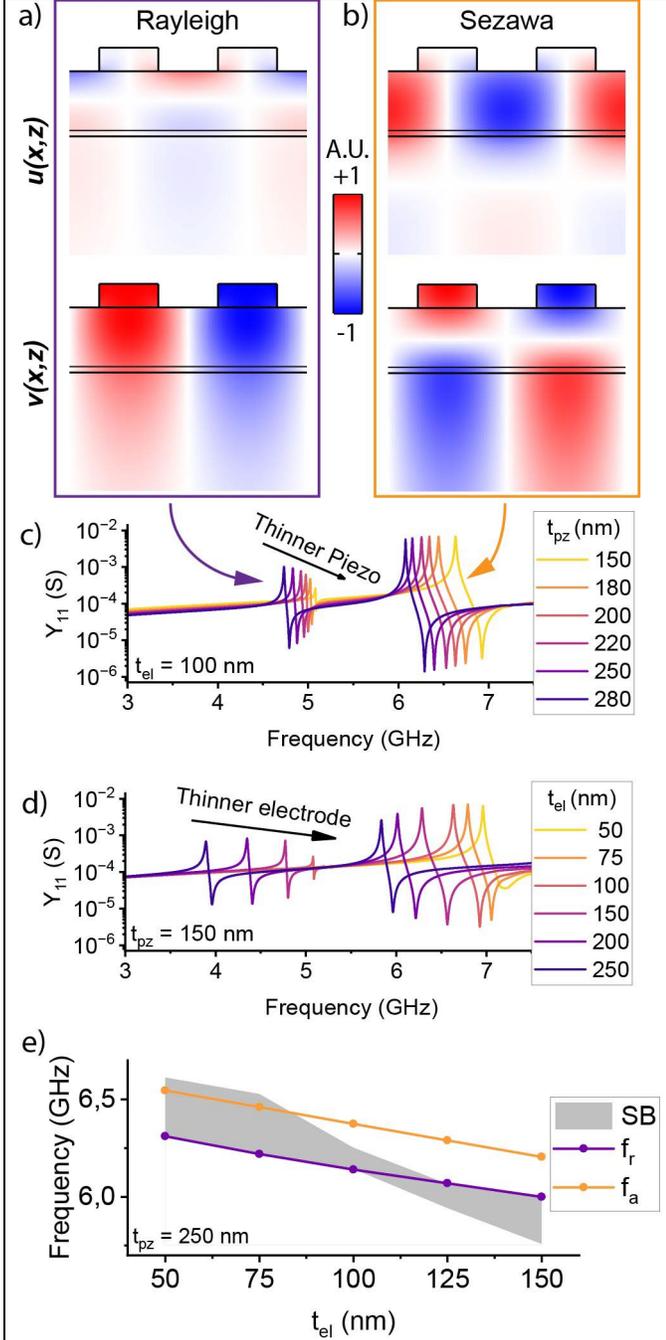

Fig. 1 a) x-axis (u) and b) z-axis (v) displacement of the two identified modes. c) and d) simulation of the suppression of the Rayleigh mode by changing the thickness of the piezoelectric layer or the top electrode, d) impact of the top electrode thickness on the stopband (SB) of the Sezawa mode.

For example, in Fig. 1.d it is visible that for the thinnest electrode of 50nm, the $f_a$ goes above the bulk shear frequency, and the energy leakage causes a visible drop in the antiresonance Q. The electrode thickness also affects the propagation of the Sezawa mode in a similar way to what is described in the Coupling-of-Modes (COM) model [16], [17], generating a stopband due to acoustic Bragg reflections in the electrode grating. Fig. 1.e shows that by changing the thickness of the top IDT, this stopband width and location can be modulated. With an electrode thickness of 75 nm or lower, the stopband fully includes the range between $f_r$ and $f_a$ and no coupled mode can appear in the resonator bandwidth. At the same time, by reducing the thickness of the electrodes, the stopband for the Rayleigh wave decreases in width. Since the stopband width is linked to the IDT internal reflection coefficient [17], the Rayleigh mode cannot be confined and leaks out.

The lower limit on the piezoelectric layer thickness depends on the nucleation quality of AlScN, which we reported working well on Si for a thickness down to 200 nm [8]. We decided to try with thicknesses of 250 nm and 150 nm. The targeted pitch was 500 nm for a wavelength of 1 um, and electrode thickness was 100 nm for the 250 nm devices and 75 nm for the 150 nm film.

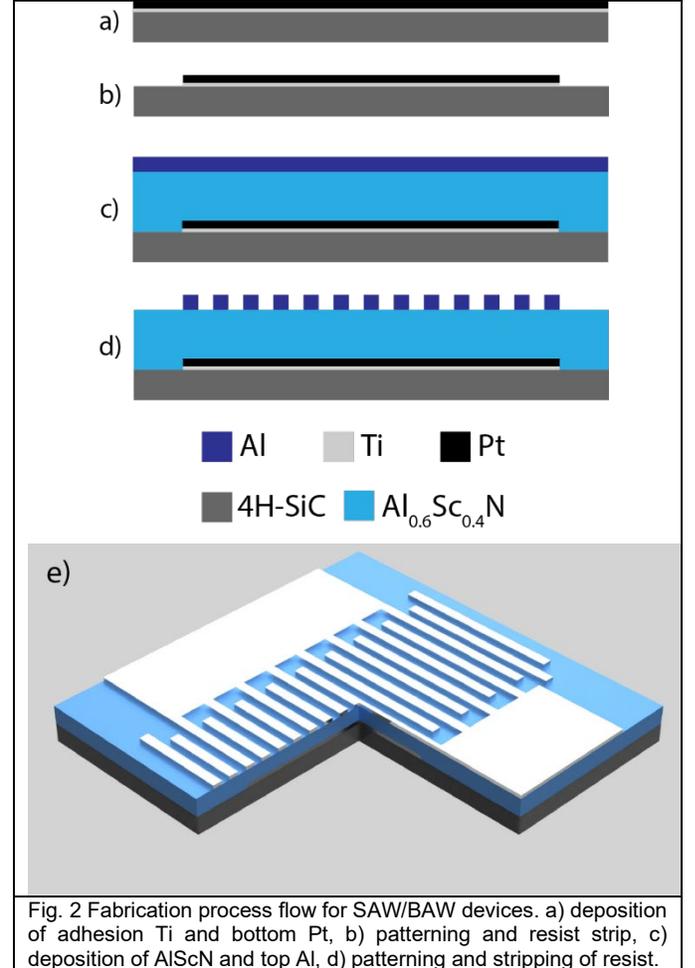

Fig. 2 Fabrication process flow for SAW/BAW devices. a) deposition of adhesion Ti and bottom Pt, b) patterning and resist strip, c) deposition of AlScN and top Al, d) patterning and stripping of resist.

### III. FABRICATION

We fabricate our wafers with a simplified process flow compared to the one of [18] since, in this case, no release is needed. We first deposit a thin layer of Ti as an adhesion layer for the Pt bottom electrode (Fig. 2.a) with the same recipe of [8]. After patterning and resist removal using a dry-wet-dry process (Fig. 2.b), we deposit the piezoelectric layer of AlScN at a temperature of 300 °C, followed by the top electrode Al layer at 100°C (Fig. 2.c). X-ray diffraction measurements show an excellent rocking curve FHWM of 1.2 ° for the 250 nm thick AlScN film and 1.6 ° for the 150 nm thick one. This, compared with the findings on Si for the same recipe in [8], shows that SiC shows a better verticality for AlScN growth than Si for a comparable thickness. The top electrodes are then patterned (Fig. 2.d) with $Cl_2$ ICP RIE, stripping the resist directly in-tool with a $CF_4/O_2$ mixture to minimize corrosion of Al from the residual $Cl_2$ [19]. Since the resonator does not require patterning of the AlScN layer, this process becomes more straightforward than previously reported[18]. The fabricated resonator, as visible in the cross-section sketch in Fig. 2.e, has a bottom floating metal electrode only below the IDT aperture to promote a vertical electric field.

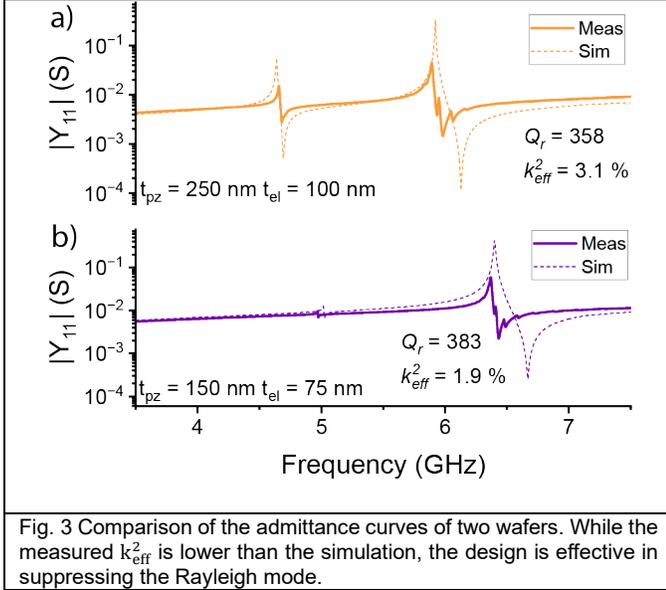

Fig. 3 Comparison of the admittance curves of two wafers. While the measured $k_{eff}^2$ is lower than the simulation, the design is effective in suppressing the Rayleigh mode.

## IV. MEASUREMENTS

After calibrating the probes using a standard SOLT procedure, the devices are measured with GSG probes and an RS-ZNB20 VNA. In Fig. 3 we plot the admittance curves of two devices with the same pitch but with different layer thicknesses compared with the simulations obtained for the same geometries. It is possible to observe how, as predicted, the Rayleigh mode disappears when $t_{pz}$ gets thinner. The $k_{eff}^2$ is lower than the simulated one. We have not found conclusive evidence for this reduction in $k_{eff}^2$, but we believe this is due to fabrication issues of the very thin layer of AlScN. Indeed, even though the film quality on large areas of bottom metal is very good (low rocking curve FWHM and low concentration of abnormally oriented grains (AOGs)), we have not been able to observe the quality on the small areas under the IDT, which comprises the active part of our fabricated resonators. Further investigation is currently ongoing to resolve this issue. It is reported in literature that in a Sezawa-mode resonator, the number of electrodes in the IDT impacts the $k_{eff}^2$ as well [14].

On the contrary, the quality factor shows acceptable values, with devices showing a loaded Q of 390 at 6.3 GHz frequency. Importantly, the layout of the top IDT is made of 142 fingers, with only 5 fingers on each side, used as dummies for the photolithography, which are too few to act as effective reflectors. This means that the wave confinement is a consequence only of the IDT grating itself, as expected from the stopband behavior shown in Fig. 1.g.

Any spurious modes, like the ones seen in the measurements of Fig. 3, originate then from transversal wavenumbers and can be suppressed by using weighted electrodes, apodization, or tilted busbars[20], [21].

## V. CONCLUSION

This letter presents a study on the design and fabrication of an AlScN-on-SiC resonator. We show how an acoustic Sezawa wave can be confined by a high phase velocity substrate like SiC and how the thickness of the piezoelectric and the electrodes help in suppressing the unwanted Rayleigh wave modes. Further investigation is necessary on the quality of these very thin films to achieve a higher coupling than the measured one. Overall, using Sezawa waves in AlScN is a promising solution to achieve high frequency, lateral mode resonators with good power handling.